\documentstyle[prd,aps,floats,psfig]{revtex}

\newcommand{\NP}[1]{ Nucl.\ Phys.\ {#1}}

\newcommand{\PL}[1]{ Phys.\ Lett.\ { #1}}

\newcommand{\AN}[1]{ Ann. Phys. {#1}}

\newcommand{\PR}[1]{Phys.\ Rev.\ { #1}}
\newcommand{\PRL}[1]{ Phys.\ Rev.\ Lett.\ { #1}}

\newcommand{\IJmp}[1]{{\em Int.\ J.\ Mod.\ Phys.\ }{#1}}

\newcommand{\AmS}{{\protect\the\textfont2
  A\kern-.1667em\lower.5ex\hbox{M}\kern-.125em}}

\newcommand{\Od}{{\cal O}}

\newcommand{\gsim}{\raise.3ex\hbox{$>$\kern-.75em\lower1ex\hbox{$\sim$}}}
\newcommand{\lsim}{\raise.3ex\hbox{$<$\kern-.75em\lower1ex\hbox{$\sim$}}}

\begin{document}
\draft
\input epsf
\renewcommand{\topfraction}{0.8}
\twocolumn[\hsize\textwidth\columnwidth\hsize\csname
@twocolumnfalse\endcsname
\preprint{hep-ph/9912512}
\title{Unitarized pion-nucleon scattering within
Heavy Baryon Chiral Perturbation Theory}

\author{A. G\'omez Nicola  and  J. R. Pel\'aez }
\address{ Departamento de F\'{\i}sica Te\'orica. \\
Universidad Complutense. 28040 Madrid. SPAIN.}
\date{February 3, 2000} 
\maketitle

\begin{abstract}
By means of the Inverse Amplitude Method 
we unitarize the elastic pion-nucleon  scattering amplitudes
obtained  from Heavy Baryon Chiral Perturbation Theory
to $\Od(q^3)$. Within this approach we can  enlarge
their applicability range  and
generate the $\Delta(1232)$ resonance.
We can find a reasonable description
of the pion nucleon phase shifts with 
$\Od (q^2)$ parameters in agreement
with the resonance saturation hypothesis. 
However, the uncertainties in the 
analysis of the low energy data as well as
the large number of chiral parameters, which
can have strong correlations, allow us to obtain
very good fits with rather different
sets of chiral constants. 
\end{abstract}
\vskip2pc]


Chiral Symmetry plays a fundamental role in the interactions between
pions and nucleons.
However, in order to go beyond current algebra 
or tree level  calculations from simple  models, 
one needs an
effective low-energy field theory with all the QCD symmetries
and a systematic power counting. 
 Heavy Baryon Chiral Perturbation Theory (HBChPT) 
\cite{HBChPT} is the best candidate to date for such 
 a theory. 
For  $SU(2)$ chiral symmetry, the HBChPT degrees of freedom
are the  nucleons and pions. The pions  are the 
Nambu-Goldstone bosons of 
the spontaneous chiral symmetry breaking,
whereas the nucleons are included as an isospin  doublet. In the
case of $SU(3)$ symmetry, the pseudoscalar meson and  baryon
octets are required to describe the
meson-baryon sector. There are also non-minimal
formulations
considering  the baryon decuplet  as a fundamental field
\cite{explicitdelta}.
  
HBChPT is  built
as an expansion in derivatives and meson masses, including all
terms compatible with chiral symmetry. It follows  the same philosophy as 
Chiral Perturbation Theory (ChPT) in the purely mesonic sector
\cite{chpt}.  However, there are
two important differences between them. 
First, the scale in the HBChPT expansion is not
just the chiral symmetry breaking scale of pion  loops 
$\Lambda_\chi=4\pi f_\pi\simeq 1.2$ GeV, but also the mass of the
nucleons  $m_B\simeq\,1\,\hbox{GeV}$. Second, the nucleon four-momentum 
is of the same order as the expansion scale, no matter how small is 
the momentum transfer and even in the chiral limit \cite{Gasser}. 
HBChPT circumvents  this problem by treating $m_B$ 
as large compared to the external momenta and redefining the nucleon fields
in terms of  velocity-dependent eigenstates
satisfying a massless Dirac equation. 
The anti-baryon components can be 
integrated out \cite{maro96}. 
 Once this is done, it is possible to find a {\em systematic} power counting 
in $k/\Lambda_\chi$, $k/m_B$, $q/\Lambda_\chi$ and $q/m_B$, where $q$
stands for any  meson mass or external momentum, generically 
denoted by $\Od(q)$. 
With the effective vertices up to  a given order,
 one can calculate loop diagrams. Each loop increases 
the order of the diagram so that any divergence 
can be absorbed by renormalising the coefficients of 
higher order operators. It is thus possible to
obtain finite results order by order for any observable, but
paying the price of introducing more  chiral parameters.
In particular,  the full HBChPT Lagrangian 
up to $\Od(q^3)$  has been given in 
\cite{EM96,Fet98}. It involves 5 unknown coefficients 
 to $\Od(q^2)$ and 23 more to $\Od(q^3)$, which
 have to be fitted to experiment. 
In this paper, we will concentrate on the pion-nucleon scattering
amplitude derived from  $SU(2)$ HBChPT 
to $\Od(q^3)$, which we will unitarize using the Inverse Amplitude
Method (IAM). 




Despite its difficulty,
 several  works related to  $\pi N$ scattering
within HBChPT
have appeared in the literature \cite{Fet98,BeKaMe,Mojzis}.  
Here we will follow the notation of the
first $\Od(q^3)$ complete result \cite{Mojzis}.  Thus,
we denote
by $a_i$ and $b_i$ the $\Od(q^2)$ and  $\Od(q^3)$ parameters,
respectively. The translation to the notation 
of \cite{Fet98,BeKaMe} can be found in \cite{Fet98}. 
 Only four  $\Od(q^2)$  and five $\Od(q^3)$    combinations
of  chiral parameters are relevant for $\pi$-N scattering. 
There are two sets of fitted parameters
in the literature:  in
\cite{Mojzis},
the {\em extrapolated} threshold values of \cite{KP80},
together with  the
nuclear $\sigma$-term and  the Goldberger-Treiman discrepancy were 
taken as the experimental input, whereas
in  \cite{Fet98}, the  $S$ and $P$-wave phase
shifts, somewhat away from threshold, were used as  data
for the fits. These  phase shifts are given in  \cite{VPI} and are
obtained as
 extrapolations from experimental
data. 
 Note that in  \cite{Buttiker}
it is suggested that the data in \cite{VPI} yield a too large
$\sigma$ term when analyzed with HBChPT. 
As their authors pointed out, both  procedures are subject to
some caveats, either because of the many uncertainties of the data near
threshold (see \cite{Fet98}),  
or because  the errors for the extrapolated phase shifts
are clearly underestimated (see \cite{Mojzis}).
Concerning theoretical estimates, in  \cite{BeKaMe}
it was suggested that the $a_i$  could be understood from
resonance exchange saturation. 
Further constraints from dispersive techniques
can be found in \cite{Buttiker}. In general, there is a fairly good
agreement between the $a_i$  values, but that is not the case 
for $b_i$ (see  \cite{Buttiker}). 

We should stress that the
predictions obtained within this framework are promising, although not 
as impressive \cite{Mojzis}
as those of ChPT for mesons. 
HBChPT is, of course, limited 
to low pion momentum and, with the presently available
calculations, certainly below  $q_\pi\lsim 200$ MeV \cite{Fet98}. 
The reason is, basically, 
that {\em the convergence of HBChPT is rather slow}. As a matter of fact,
the contributions of the first three orders are frequently comparable.

In order to improve this situation, one could go to 
the next order, thus dealing  with 
many more parameters. We could also  introduce more  
degrees of freedom, like the lightest resonances \cite{others},
but that would also increase the number of parameters.
In addition, some
kind of unitarization should also be carried out in that case, 
to impose strict unitarity.
Recently \cite{JA}, remarkable results have been obtained
by unitarizing with the  N/D method the
lowest order HBChPT at tree level, including explicitly the 
$\Delta(1332)$ and $N^*(1440)$
resonances.

We propose an alternative approach.
Encouraged by the HBChPT results
and the success of unitarization 
in meson-meson scattering \cite{IAM,oop}, we will
unitarize the amplitude 
\emph{without introducing additional fields}.




\subsection{The IAM applied to $\pi$-N scattering.}
\label{iam}

Unitarization is not 
foreign to effective theories. In fact,  Pad\'e approximants 
with very simple  models are
enough to describe the main features of $\pi$-N scattering
 \cite{70s}. Although 
a systematic application within an effective Lagrangian
approach was called for, it was  never  carried out. 

Customarily \cite{ericson}, 
the data are 
presented in terms of 
partial waves of definite isospin $I$, orbital angular momentum $L$
and total angular momentum $J$, using
the spectroscopic notation $L_{2I+1,2J+1}$ (with
$L=S,P$,...waves). Generically,
the HBChPT $\pi$-N partial waves, $t$,
are obtained as a series in the momentum transfer and meson masses.
Thus, basically, they are  
polynomials in the energy and mass variables (as well as logarithms
from the loops, which provide the cuts and imaginary
parts required by unitarity). Such an 
expansion will never satisfy 
the $\pi$-N {\em elastic} unitarity condition 
\begin{equation}
  \hbox{Im}\, t = q_{cm}\, \vert t\vert^2 \Rightarrow 
\hbox{Im} \frac{1}{t} = -q_{cm}  \Rightarrow 
 \frac{1}{t}=\hbox{Re} \frac{1}{t}-i\, q_{cm}, 
\label{unit}
\end{equation}
with $q_{cm}$ the center of mass
momentum of the incoming pion. But 
HBChPT satisfies unitarity 
{\em perturbatively}, i.e.:
  \begin{eqnarray}
  \hbox{Im}\, t_1 = \hbox{Im}\, t_2 = 0\;;\;
  \hbox{Im}\, t_3 = q_{cm}\,\vert  t_1\vert^2, \dots, 
\label{pertunit}
\end{eqnarray}
where $t_k$ stands for the $\Od(q^k)$ contribution to the amplitude. 
This is indeed the case in \cite{Fet98}, but not  in \cite{Mojzis},
where an additional 
redefinition of the nucleon field allows to 
eliminate the $(v\cdot\nabla)^2/2m$ terms 
in the Lagrangian \cite{EM96}.
 We have performed  an additional $1/m$ expansion
of the results in  \cite{Mojzis} in order to 
 recover a pure expansion satisfying
eq.(\ref{pertunit}). Thus, we  count  
 $q_{cm}$ and $M$ as $\Od(\epsilon)$, so that 
each partial wave reads
 $t\simeq t_1+t_2+t_3+ \Od(\epsilon^4)$, 
where the subscript stands for the order $\epsilon$ of
each contribution.   We have then checked that
eq.(\ref{pertunit}) is verified.

However, from eq.(\ref{unit}) any unitarity elastic amplitude
has  {\em exactly} the following form:
\begin{equation}
t=\frac{1}{\hbox{Re}(1/t)-i\, q_{cm}}, 
\label{tunitaria}
\end{equation}
\emph{for physical values of the energy, and 
below any inelastic threshold}.
The problem, of course, is how to obtain
Re(1/t). For instance, setting 
$\hbox{Re}(1/t)= \vert q_{cm}\vert\, \hbox{cot}\delta
=-\frac{1}{a}+\frac{r_0}{2}\, q_{cm}^2$ we reobtain the familiar
effective range approximation, whereas by taking 
$\hbox{Re}\, t\simeq t_1$
 we arrive at a Lippmann-Schwinger like  equation \cite{oop}. 
A frequent criticism to unitarization is 
its apparent arbitrariness, although from 
eq.(\ref{tunitaria}) we  see that the difference between two
 unitarization methods is the way of approximating Re$(1/t)$. 
Since we want to restrict our Lagrangian to include just pions
and nucleons preserving  the $SU(2)$ chiral symmetry, the most general
approximation to Re(1/t) is HBChPT. In particular,
we will take the $\Od(q^3)$ calculations, 
but the method can be easily generalized to higher orders.
Thus, we arrive at
\begin{equation}
t\simeq \frac{t_1^2}{t_1-t_2+t_2^2/t_1-\hbox{Re}\,t_3-iq_{cm}t_1^2},
\label{IAM3}
\end{equation}
where we have only kept the relevant order in
Re$[(t_1+t_2+t_3)^{-1}]$. 
 This is  the $\Od(q^3)$ form of the IAM. 
Note that if
we reexpand in powers of $q$, we recover at 
low energies the HBChPT result.
However, as it is written,
the amplitude explicitly satisfies elastic unitarity.
Furthermore, using eq.(\ref{pertunit}), 
we can rewrite $\hbox{Re}\,t_3+iq_{cm}t_1^2=t_3$,
which can be analytically continued
to the complex plane (where, for instance, we will look for ç
the pole associated to the $\Delta(1232)$).
Incidentally, eq.(\ref{IAM3}) thus rewritten is 
a Pad\'e approximant of the $\Od(q^3)$ series.
From the K matrix point of view, we have identified
$K^{-1}=\hbox{Re}\, t^{-1}$.

Even though the elastic unitarity condition is only
satisfied for real values of $s$ above threshold, the 
use of the IAM in the complex plane 
can be justified using dispersion 
relations \cite{IAM},
provided one is not very far from the physical cut.
In other regions (around the left cut for instance)
the IAM would be inappropriate. 
As a consequence, it is also
possible to reproduce the poles in the second Riemann sheet, which are 
close to the physical cut and are
associated to resonances. As a matter of fact
the IAM has been successfully applied to meson-meson scattering
\cite{IAM}. In particular, 
using the $\Od(p^4)$ ChPT Lagrangian, the IAM generalized to 
coupled channels yields a remarkable description of
all channels up to 1.2 GeV, including seven resonances \cite{oop}.
Using the Lippmann-Schwinger like  equation mentioned
above it is also possible to describe the S-wave kaon-nucleon scattering, 
using the lowest order Lagrangian \cite{or}, including
the $\Lambda(1405)$.

Let us then use eq.(\ref{IAM3}) when $t$ 
are the $L_{2I+1,2J+1}$ $\pi$-N partial waves. 
The resulting amplitudes will be fitted to the
\cite{VPI} phase shifts, which are actually an
extrapolation, not including the experimental errors. 
For the fit we have used the MINUIT Function
Minimization and Error Analysis routine from the CERN program Library.
As it is customarily done in the literature, we will 
assign an error to the data in \cite{VPI}.
For instance, in ref.\cite{Fet98} the central points
have been given a 3\% uncertainty. However, since our fits will cover
wide energy ranges, the use
of a constant relative error will give more
weight to the low energy data. Thus
we have also added an additional systematic error of 1 degree.
( A 5\% error plus a $\sqrt{2}$
systematic error was used in \cite{JA}.)
This error is needed to use the
minimization routine, and,
although the order of magnitude may seem appropriate, 
the values are rather arbitrary, so that
the meaning of the $\chi^2/d.o.f.$ obtained from 
MINUIT has to be interpreted cautiously.

Furthermore, the data near threshold are subject to many
uncertainties, so that, also following \cite{Fet98}, 
we will start our fits at $\sqrt{s}=$1130 MeV. Hence
the threshold parameters are real predictions in our approach.  
In addition, we should limit the approach to 
energies where inelasticities can be neglected.
In particular, we will not use our 
$P_{11}$, $P_{13}$ and $P_{31}$ phase shifts above
the $\pi\pi N$ threshold ($\sqrt{s}\simeq$ 1220 MeV),
since they are very small and 
inelasticities could be significant.
The $S_{31}$ and $S_{11}$ phase shifts are larger and 
we fitted them up to  $\simeq$ 1360 MeV.
The inelasticities in $P_{33}$ are negligible
up to 1400 MeV since this channel is dominated by the $\Delta(1232)$
which is strongly coupled to $\pi$-N.

\begin{figure}
\begin{center}
\vspace*{-.6cm}
\hbox{\psfig{file=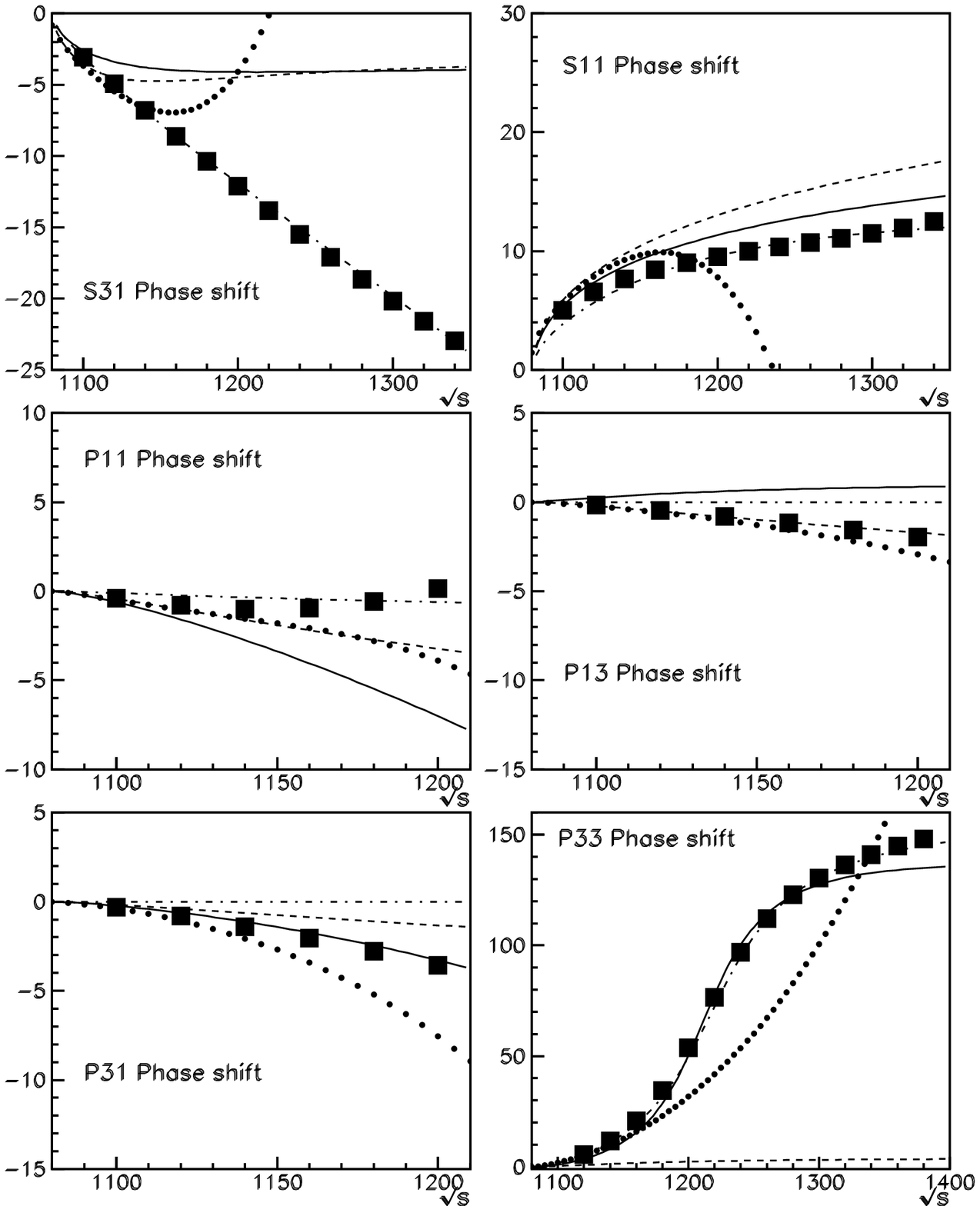,width=8.cm}}  
\end{center}
\vspace*{-.7cm}
\caption{\label{fit1}}
{\footnotesize 
$\pi$-N scattering phase shifts.
The dotted curve is the extrapolated HBChPT result, 
with the chiral parameters of \cite{Mojzis}.
The dashed line is the IAM
with the same parameters, and the continuous line 
is the IAM constrained to resonance saturation
(fit 2, see text). The dashed-dotted line is one unconstrained IAM
fit. The data come from \cite{VPI}.}

\end{figure}

\paragraph{The IAM and Resonance Saturation.}
Following the suggestion that the $O(p^2)$ parameters can
be understood from resonance saturation \cite{BeKaMe},
it is natural to try to make an IAM fit constrained
with this hypothesis.  Thus, we first
fix the $a_i$ to the values of \cite{Mojzis},
which are  compatible with the saturation hypothesis.
The resulting  $O(p^3)$ parameters are given in 
Table \ref{tab2}. 
In addition, we give in Table \ref{tab2} the values for 
a second fit where we have
allowed the $a_i$ parameters to vary within the ranges
expected from resonance saturation. The results of
fit 2 are plotted, as a solid line, in Figure \ref{fit1}.
We used ``hatted'' quantities, $\hat b_i$,
because their values do not necessarily correspond to those of
HBChPT since now they are also absorbing 
the IAM resummation effects and some high energy information.
Only if there was a very good convergence of the
theory at low energies the values of $b_i$
should be similar to the $\hat b_i$ (as it happens
in ChPT). Therefore, at present, \emph{our  $\hat b_i$ should
not be used to calculate any other process at low-energies}.

Not surprisingly, there are
strong correlations between parameters. Unfortunately, 
from MINUIT we cannot get the actual form of the correlation. 
However, looking only at linear combinations
with integer coefficients, some of them, 
like $\hat b_1+\hat b_2+\hat b_3$ or
$\hat b_1+\hat b_2+2\hat b_3+\hat b_{16}-\hat b_{15}$, 
remain within natural sizes for these fits.
Nevertheless, as we have commented, there is a considerable
uncertainty in the precise values of the $b_i$ at present
(see Table 2 in \cite{Buttiker}). 

In summary, the main conclusion
from Figure \ref{fit1} is that it is possible to obtain
an improved description of $\pi N$ scattering including
the $\Delta(1232)$, with the $a_i$ values
obtained from resonance saturation. 
Note however, that the $S_{31}$ phase shift does not have
any real improvement.
For illustration  we also give
in Figure \ref{fit1} the extrapolation of the $O(q^3)$ HBChPT results 
to high energies
(dotted line) as well as the IAM result (dashed line),
using the  $a_i$ and $b_i$ values in \cite{Mojzis}.

\paragraph{Unconstrained IAM fits.}
Of course, we can get much better fits (all with $\chi^2/d.o.f.\lsim
1$) by leaving all 
the parameters free. For illustration, see the dashed-dotted line in
Figure \ref{fit1}. 
There are again strong correlations and
the actual value of each one of the
$\hat a_i$ and $\hat b_i$ could be extremely unnatural.
The correlations now are even more complicated due
to the quadratic $t_2^2$ term in the denominator of eq.(\ref{IAM3}).
By inspection of the analytic formulas, we find that
the $\hat a_1+\hat a_2$, $\hat a_5-4\hat a_1$, 
$2(\hat b_1+\hat b_2)+(\hat b_{16}-\hat b_{15})$
combinations  are the most relevant, and remain rather stable
for these fits. However\emph{ it is not possible to obtain a meaningful 
determination of each individual
parameter} without any other additional assumption
(like resonance saturation). That is again
due to the large number of parameters, but also
to the slow HBChPT convergence.

\paragraph{The $\Delta(1232)$ resonance.}
The IAM generates dinamically a pole in the second Riemann
sheet at $\sqrt{s}=(1212-i\,47)\,\hbox{MeV}$, which 
is rather stable within all the fits and in 
very good agreement with the data \cite{PDG}.


\paragraph{Threshold parameters.}
For definitions and notation
we refer again to \cite{Mojzis}. Our  results 
are shown in Table \ref{tab3}, where we have also listed the
experimental values, extracted from \cite{KP80}.   As pointed out
in \cite{KP80} and 
\cite{Mojzis}, the errors for those values  are clearly {\it underestimated}. 
Hence, it  should be borne in mind that
the threshold parameters  are not so well determined as it may seem from
those errors (see also \cite{Fet98}). 
Our fits give a reasonably  good agreement with experiment for the $S$-wave
scattering lengths. For most $P$-waves, 
 we agree with the order of magnitude and sign.
Our results are  also in rough agreement with
\cite{Fet98}, where they give: -0.07 GeV$^{-1}$ $\leq a_0^+\leq$ 0.04
GeV$^{-1}$ and   0.6 GeV$^{-1}$ $\leq a_0^-\leq$ 0.67
GeV$^{-1}$.

\subsection{Conclusions and discussion}
\label{conc}

We have unitarized
the  HBChPT $\Od(q^3)$ $\pi$N elastic scattering amplitude
with the Inverse Amplitude Method.
This approach is able to describe
 the phase shifts up to the inelastic thresholds and, 
in addition, it gives the
correct pole for the $\Delta (1232)$ in the $P_{33}$ channel. 
Our fits use the extrapolated phase shifts between 
$\sqrt{s}=$ 1130 MeV and the corresponding
inelastic thresholds. Within this approach, we can predict the
thresholds values, which for the $S$ waves are in good
agreement with experiment and with recent determinations. 

Unfortunately, since there are
large correlations between some parameters,
it is possible to obtain good fits with very
different sets of parameters, which can have rather unphysical values. 
This is due to different reasons: a) The slow convergence of the series,
since contributions from different orders are
comparable in almost every partial wave. 
The effect of higher order terms, which was less relevant at threshold,
is absorbed in our case in the values of the chiral coefficients.
b) There are strong correlations between the parameters,
and the fits are only sensitive to certain combinations. Hence the
values of each individual coefficient are meaningless. 

The most relevant conclusion of this study is that
we can still reproduce the $\Delta (1232)$ with the 
$a_i$ values expected from the
resonance saturation hypothesis, keeping a  reasonably good
description for the other channels.

Finally, we would like to remark that the method developed here can be 
easily extended to the case of $SU(3)$ symmetry as well as to
the coupled channel formalism. Further work along
these lines is in progress.

\section*{Acknowledgments}
Work partially supported by DGICYT under contract
AEN97-1693 and PB98-0782. 
J.R.P. thanks J.A.Oller and E.Oset for useful discussions.

\begin{table}
\caption{\label{tab2}}
{\footnotesize 
IAM results. In fit 1, we keep the 
$\Od(q^2)$ values of \cite{Mojzis}.
In fit 2, the $a_i$ are constrained
 to the ranges predicted by
resonance saturation \cite{BeKaMe}.
Due to the  strong correlations (see text)
only certain $\hat b_i$ combinations could
be meaningful.}
   \setlength{\tabcolsep}{1.mm}
\begin{tabular}{|c||c|c|c|c||c|c|c|c|c|}
\hline
&$a_1$ & $a_2$ & $a_3$ & $a_5$ & $\hat b_1+\hat b_2$&  
$\hat b_3$&$\hat b_6$&$\hat b_{16}-\hat b_{15}$&$\hat b_{19}$\\  \hline
Fit 1
&-2.6 &1.4 &-1.0&3.3&28.1&-29.8&2.1&33.3&12.9 \\\hline
Fit 2
&-2.1&1.3 &-0.8&3.6&22.3&-26.1&2.5&26.7&10.1 \\\hline
\end{tabular}
\end{table}

\begin{table}
\caption{\label{tab3}}
{\footnotesize 
$\pi$N threshold values with different IAM fits, 
and their experimental values (from \cite{KP80}, see text).}
\vspace*{-.3cm}
\begin{center}
\begin{tabular}{|c|c|c|c|}
\hline
 \hspace*{.5cm}&{\footnotesize
\begin{tabular}{c}
Extrapolated\\ 
from \\Experiment
\end{tabular}}&
{\footnotesize
\begin{tabular}{c}
IAM constrained \\
to resonance\\ 
saturation (fit 2)
\end{tabular}}
&
{\footnotesize
\begin{tabular}{c}
IAM \\
unconstrained 
\end{tabular}}
\\\hline
$a_0^+$ (GeV$^{-1}$)& -0.07$\pm$ 0.01&   0.02 &-0.12 \\ \hline
 $a_0^-$ (GeV$^{-1}$)&0.67$\pm$ 0.1&     0.72 &0.53 \\ \hline
$b_0^+$ (GeV$^{-3}$)&-16.9$\pm$2.5&       -2.8&-16.43 \\ \hline
 $b_0^-$ (GeV$^{-3}$)&5.1$\pm$2.3&        -9.0&13.59  \\ \hline
 $a_{1+}^+$ (GeV$^{-3}$)&50.5$\pm$0.5&   41.04&79.08 \\ \hline
 $a_{1-}^+$(GeV$^{-3}$)&-21.6$\pm$0.5&   -18.35&-2.09 \\ \hline
 $a_{1+}^-$ (GeV$^{-3}$)&-31.0$\pm$0.6&  -11.52&-39.58 \\ \hline
$a_{1-}^-$(GeV$^{-3}$)&-4.4$\pm$0.4&      -6.75&-2.00 \\ \hline
\end{tabular}
\end{center}
\end{table}

\end{document}